\documentstyle[times,pramana,epsf,floats]{ias}

\newcommand{\tauiso}{{\mbox{\boldmath $\tau$}}}
\begin{document}
\mark{}
\title{Dilepton production in nucleon-nucleon collisions 
around 1 GeV/nucleon: a theoretical update}

\author{R. Shyam$^{1,2}$, and U. Mosel$^1$}
\address{$^1$ Institut f\"ur Theoretische Physik, Universit\"at Giessen,  
D-35392 Giessen, Germany \\
$^2$ Saha Institute of Nuclear Physics, 1/AF Bidhan Nagar, Kolkata 700064, 
India}
\keywords{Dilepton production, NN collisions, effective Lagrangian model}
\pacs{25.75.Dw, 13.30.Ce, 12.40.Yx}
\abstract{We present a fully relativistic and gauge invariant framework for
calculating the cross sections of dilepton production in nucleon-nucleon
($NN$) collisions which is based on the meson-exchange approximation for
the $NN$ scattering amplitudes. Predictions of our model are compared with
those of  other covariant models that have been used to describe
this reaction.  We discuss the comparison of our calculations with the old DLS
and the recent HADES data.
}

\maketitle
\section{Introduction}
Experiments with High Acceptance Di-electron Spectrometer (HADES) are
aimed at searching for medium modifications of hadrons at moderately high
temperatures ($T < 100$ MeV) and baryonic densities up to 3 times the
normal nuclear matter density. Due to negligible final state interactions
with surrounding medium, dileptons ($e^+e^-$) provide a very clean and 
powerful probe for this purpose.

A recurring feature of the dilepton  spectra measured in nucleus-nucleus 
($AA$) collisions has been the enhancement (above known sources) in the 
invariant mass distribution of the cross sections in the intermediate 
region of dilepton invariant masses ($M$). This has been the case for 
experiments performed for bombarding energies ranging from as low as 
1.0 GeV/nucleon (DLS and HADES data~\cite{por97,aga09}), through the 
SPS energies (40 - 158 GeV/nucleon)~\cite{aga05,ada03,arn06} to the energies 
employed by the PHENIX collaboration at RHIC (which correspond to invariant 
mass of 200 GeV/nucleon)~\cite{afa07}. The enhancement seen at the SPS energies 
are understood in terms of the modification of the $\rho$ meson spectral 
function in the hadronic medium \cite{rap07}.

The large dileptons yields observed in the DLS experiment (even
for the light collision systems $^{12}$C + $^{12}$C) in the invariant mass 
range of 0.2 - 0.5 GeV, are yet to be explained satisfactorily
\cite{ern98,cas99,rap00,she03,aga07}. Independent transport model calculations 
(TMC) have been unable to describe these data fully even after including 
contributions from (i) the decay of $\rho$ and $\omega$ mesons which are 
produced directly from the nucleon-nucleon ($NN$) and pion-nucleon scattering 
in the early reaction phase~\cite{bra98}, (ii) the in-medium $\rho$ spectral 
functions \cite{cas98}, (iii) the dropping $\rho$ mass with corresponding 
modification in the resonance properties~\cite{ern98}, (iv) an alternative 
scenario of the in-medium effect - a possible decoherence between the 
intermediate meson states in the vector resonance decay \cite{she03}. This 
led to term this discrepancy as "DLS-puzzle" \cite{ern98,bra98} which persists 
even now. The new dilepton production data of the HADES Collaboration agree 
remarkably well with the corresponding DLS data~\cite{aga08}. Therefore, there 
is no longer any question against the validity of the DLS data and the  
failures to explain them by various transport models have to do with problems 
in the theoretical calculations. 

Cross sections for dilepton production in elementary nucleon-nucleon ($NN$) 
collisions are important inputs in the TMC. For the $NN$ bremsstrahlung 
processes most TMC use the cross sections predicted by a semiclassical soft photon 
approximation (SPA) model~\cite{ruc76,gal87}. An important point made in a 
recent TMC~\cite{bra08}, is that if one scales up the SPA cross sections for 
the  $pp$ and $pn$ systems by factors of 2-4, the observed dilepton yields of 
both DLS and HADES experiments can be reproduced. These enhancement factors 
are motivated by the larger elementary $NN$ dilepton cross sections predicted 
by the calculations reported in Refs.~\cite{kap06,kap09}.  However, these large 
cross sections are not in agreement with those reported in several previous 
studies \cite{sch94,dej96,shy03}. They also do not agree with results of a 
more recent study of the dilepton production in $NN$ collisions~\cite{shy09}. 
Therefore, the puzzle seems to have reduced to a proper understanding of the 
dilepton production in elementary $NN$ collisions. 

In the next section we present a brief comparison of various models used to 
calculate the dilepton yields in elementary $NN$ collisions. It is 
important to understand the differences seen in the predictions of 
various models in order to have a proper theoretical description of the new 
HADES data on the dilepton production not only in elementary $pp$ and $pn$ 
collisions but also in the $AA$ collisions.

\section{Comparison of various models of dilepton production in
elementary $NN$ Collisions}

Calculations reported in Refs.~\cite{kap06,kap09} (to be referred as KK),
and~\cite{shy03} (to be referred as SM1) use the same basic model to calculate 
the dilepton production in $NN$ collisions. Both KK and SM1 account for the 
initial interaction between two incoming nucleons by an effective Lagrangian 
which is based on the exchange of the $\pi$, $\rho$, $\omega$ and $\sigma$ 
mesons. The coupling constants and the form factors at the nucleon-nucleon-meson 
vertices are the same in both studies. These parameters were determined in 
Ref.~\cite{sch94} by directly fitting the $T$ matrices of the $NN$ scattering 
in the relevant energy region, and they have been used in the 
successful descriptions of the $NN \to NN\pi$~\cite{shy98}, 
$pp \to p\Lambda K^+$, $pp \to p\Sigma^0 K^+$~\cite{shy99,shy01} and
$NN \to NN\eta$~\cite{shy07} reactions. 

The major difference between calculations of Refs.~\cite{kap06} and \cite{shy03}
lies in the method of implementing the gauge invariance of the $NN$
bremsstrahlung amplitudes. To investigate this issue, the dilepton production
cross sections within the SM1 model have been recalculated recently~\cite{shy09}
(to be referred as SM2) by using a pseudoscalar (PS) coupling 
\begin{eqnarray}
{\cal L}_{NN\pi} & = & ig_{NN\pi} {\bar{\Psi}}_N \gamma _5
                             \tauiso \cdot {\bf \Phi}_\pi \Psi _N,
\end{eqnarray}
\noindent
for the nucleon-nucleon-pion ($NN\pi$) vertex while the KK model 
employs a pseudovector (PV) coupling 
\begin{eqnarray}
{\cal L}_{NN\pi} & = & -\frac{g_{NN\pi}}{2m_N} {\bar{\Psi}}_N \gamma _5
                             {\gamma}_{\mu} \tauiso
                            \cdot (\partial ^\mu {\bf \Phi}_\pi) \Psi _N. 
\end{eqnarray}
for this vertex. With a PV vertex an extra term (known as contact or seagull term) 
appears in the model which comes from the electromagnetic coupling (by the 
\begin{figure}
\epsfxsize=4.5cm
\begin{center}
\begin{tabular}{cc}
\epsfbox{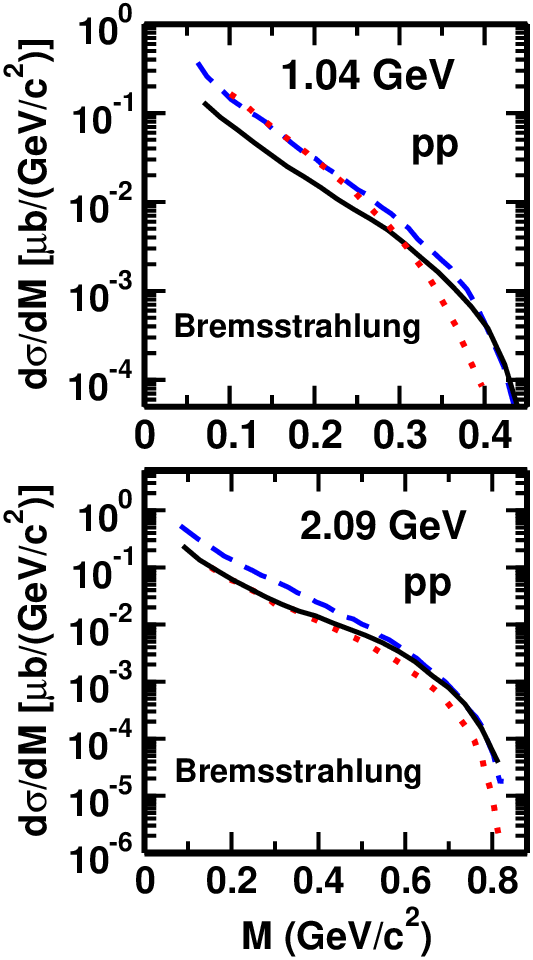} & \hspace{0.99cm}
\epsfxsize=4cm
\epsfbox{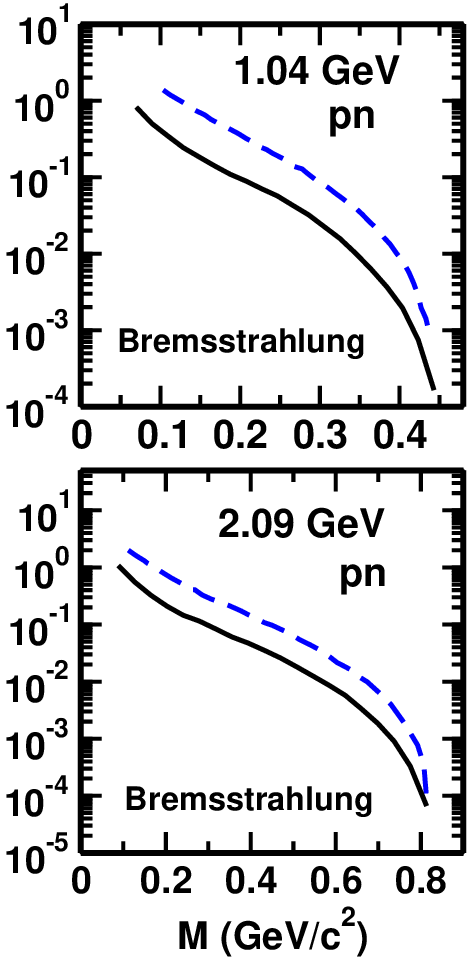}
\end{tabular}
\end{center}
\caption{The invariant mass distribution of the $NN$ bremsstrahlung contributions 
to the dilepton spectra in $pp$ (left) and $pn$ (right) collisions at the beam
energies of 1.04 GeV and 2.09 GeV. Results obtained within our model are
shown by solid lines while those of Refs.~\protect\cite{dej96} and
\protect\cite{kap06} by dotted and dashed lines, respectively.
}
\label{fig:brems}
\end{figure}
\noindent
substitution $\partial^\mu \to \partial^\mu - iem A^\mu$, where $m$ is +1, 0, 
-1 for positive, neutral and negative pions) in the $NN\pi$ 
Lagrangian. On the other hand, with a PS vertex, the theory is free from 
this term. For  point-like particles both prescriptions lead to gauge invariant 
$NN$ bremsstrahlung amplitudes. However, nucleons have finite sizes -  to 
account for this fact form factors are introduced at the corresponding vertices.
Form factors are also required to quench the unphysical contributions from   
higher momenta. In the presence of form factors the gauge invariance may be 
violated. As is shown in Ref.~\cite{sch94}, in case of the PS coupling it is 
straight forward to introduce the form factors while retaining the gauge 
invariance by following the method of Ref.~\cite{ris84}. However, with contact 
terms present in the theory in the PV case, the introduction of form factors 
without violating the gauge invariance, is a more delicate issue. There are 
several prescriptions for this and various procedures may lead to quite 
different results (see, e.g.,~\cite{uso05}. Therefore, the use of the PS 
coupling for the $NN\pi$ vertex makes the dominant pion exchange contributions 
free from such ambiguities.

In Ref.~\cite{dej96} (to be referred as dJM model), instead of the one-boson 
exchange picture of SM1, SM2 and KK calculations, the nucleon-nucleon interaction 
is included via $T$-matrices that are based on the Paris potential. This model 
provides a reliable description of the off-shell behavior of the effective $NN$ 
interaction. In addition, the rescattering terms are included in both nucleon 
and $\Delta$ intermediate states - these terms are not present in both SM (1 $\&$ 2) 
and KK calculations. Furthermore, the degeneracy of the PS and PV choices for the 
pion coupling is resolved to a large extent in this model; one has to simply use 
different parameters to obtain the same fit. However, the price paid here is that 
the nucleon current is not gauge invariant. These authors rectify this problem in 
an ad-hoc manner.

In the next section we present a comparison of the predictions of various models
for the invariant mass distributions of the dileptons produced in $NN$ collisions.

\section{Comparison of cross sections calculated within various model}
\begin{figure}
\epsfxsize=6cm
\begin{center}
\epsfbox{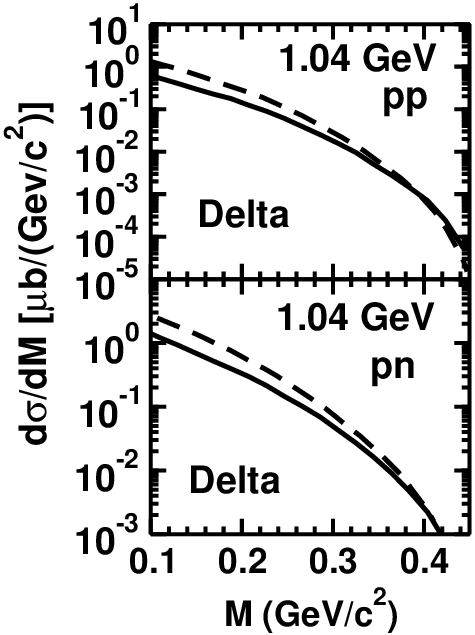} 
\end{center}
\caption{The invariant mass distribution of the $\Delta$ isobar contribution to the
dilepton spectra in $pp$ (upper panel) and $pn$ (lower panel) collisions
the beam energy of 1.04 GeV. The results of our model are shown by full lines
while those of Ref.~\protect\cite{kap06} by dashed line.
}
\label{fig:del}
\end{figure}

In Fig.~1, we show the invariant mass distribution of the $pp$ and $pn$
bremsstrahlung contributions to the dilepton spectra at beam energies of
1.04 and 2.09 GeV as calculated in SM2 model (solid lines). Also shown here 
are the results for this reaction as reported in Refs.~\cite{dej96,kap06}.
We first discuss the results for the $pp$ reaction shown in the left panel of 
this figure.  At the outset we remark that the SM2 cross sections are very 
similar to those reported in Refs.~\cite{sch94} and~\cite{shy03}. However, some 
differences are seen between the cross sections of SM2 and dJM models and also
between them and the KK calculations. Whereas the dJM cross sections are larger 
than those of SM2 in the mass region $M$ $<$ 0.25 GeV, they are almost identical 
to the latter at 2.09 GeV in this mass region. At both the beam energies, dJM 
results are smaller than SM2 ones for $M >$ 0.25 GeV. In contrast to this, the 
KK cross sections are larger than those of SM2 everywhere for $M < 0.6$ GeV. An 
important point to note is that there is no overall multiplicative factor that 
differentiates the results of various models.

It is rather surprising that despite using the same diagrams, input parameters 
and gauge invariance restoration procedure, the SM2 $pp$ bremsstrahlung cross 
sections are lower than those of Ref.~\cite{kap06}. Of course, in 
Ref.~\cite{kap06} a pseudovector $NN\pi$ vertex has been used
as compared to the pseudoscalar one employed in SM2. In this context,
it is worthwhile to note that for the real photon production, the covariant
model calculations do not depend on the choice of the $NN\pi$ coupling
(PS or PV) as is shown in Ref.~\cite{sch91}. In case of dileptons, different
results can arise for two couplings from the magnetic part of the
$NN\gamma$ vertex. In fact, in Ref.~\cite{dej97} it is shown that $pp$
dilepton bremsstrahlung contributions obtained with the PV $NN\pi$ coupling
are actually smaller than those calculated with the PS one at the beam
energy of 2 GeV. The calculations presented in Ref.~\cite{shy03} also
support this to some extent. 

In the right panel of Fig.~1, we compare the SM2 results with those of the KK 
model for the $pn$ collisions. In this case, the situation is even more 
contrasting - the SM2 cross sections are a factor of 3-4 lower than the KK ones. 
The results of the dJM model are not available for this reaction. It is  
highly desirable that this discrepancy between the results of the two models
for the $pn$ case is properly understood. 

In Fig.~2, we show a comparison of SM2 and KK results for the invariant mass 
distribution of the $\Delta$ isobar contribution to the dilepton production 
in $pp$ and $pn$ collisions at the beam energy of 1.04 GeV. We note that 
here too the KK cross sections are larger than SM2 by factors of $\sim$ 2 
at smaller values of M even though the two models have used the same 
ingredients and input parameters for this part and there is no ambiguity 
related to gauge invariance as the resonance vertex is gauge invariant by 
its very construction.
\begin{figure}
\epsfxsize=7cm
\begin{center}
\epsfbox{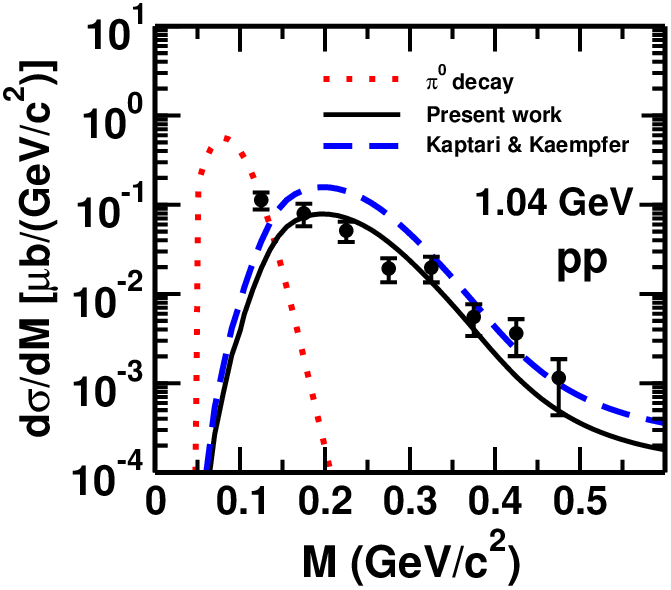} 
\end{center}
\caption{The calculated dilepton invariant mass distribution for the $pp$ 
collision at the beam energy of 1.04 GeV in comparison to the DLS data. 
The contribution of the $\pi^0$ Dalitz decay is also shown here which is the 
same as that in Ref.~\protect\cite{shy03}.
}
\label{fig:exp}
\end{figure}

In Fig.~3, we compare the SM2 and KK total cross sections for the dilepton 
production in $pp$ collisions with the DLS data at the beam energy of 
1.04 GeV. The cross sections calculated within SM2 model are folded with 
appropriate experimental filter and final mass resolution. The folded KK 
cross sections have been obtained by assuming that the folding procedure 
does not affect the ratios of the unfolded cross sections in the two cases. 
In this figure we have also shown cross sections for the $\pi^0$ Dalitz 
decay ($\pi^0 \rightarrow \gamma e^+e^-$) which are the same as those  
in Ref.~\cite{shy03}. It is seen that KK cross sections overestimate the DLS
data for $M <$ 0.3 GeV where statistical errors are smaller. The data have
larger error bars for $M >$ 0.3 GeV. In this context, the HADES data on the
elementary dilepton production reactions are useful because of their low 
statistical error. A comparison of the HADES dilepton mass spectra measured 
in the $pp$ and quasi-free $np$ reactions~\cite{aga09a} and the predictions of 
the SM2 and KK models, has been reported in Ref.~\cite{ram09}. There one 
notices that for the case of $pp$ reaction, once again the  predictions of 
the SM2 model are in pretty good agreement with the data while the KK model 
overestimates them.

In Fig.~4, we examine the role of various meson exchange processes to the 
dilepton production in the $pn$ collisions at 1.25 GeV beam energy. In this 
figure, the dotted and dashed lines show the  individual contributions of 
the $\pi$- and $\rho$-exchange graphs to the dilepton invariant mass 
distribution. Contributions of other meson exchange processes are relatively 
smaller. The coherent sum of the two terms is shown by the solid line. It 
is clear that the $\pi$-exchange process dominates the cross sections 
and more significantly, it interferes destructively with the $\rho$-exchange 
term as the coherent sum of the two diagrams is even smaller than the 
pion-only exchange term. This implies that a larger $\rho$-exchange term 
\cite{kap09} will lead to even smaller total cross sections.

\begin{figure}
\epsfxsize=6cm
\begin{center}
\epsfbox{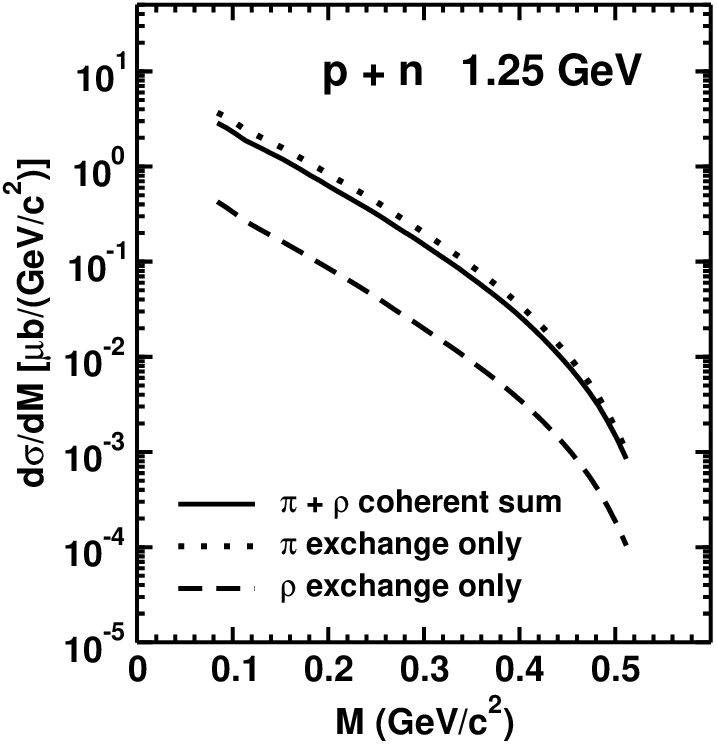}
\end{center}
\caption{
Contributions of $\pi$ (dotted line) and $\rho$ (dashed line)
exchange diagrams to the invariant mass distribution of the dileptons
produced in proton-neutron collisions at the beam energy of 1.25 GeV.
Their coherent sum is shown by the solid line.
}
\label{fig:mesex}
\end{figure}
\subsection{Calculations at the HADES energies}

In the experiments carried out by the HADES Collaboration, proton and 
deuteron beams with kinetic energies of 1.25 GeV/nucleon were incident 
on a proton target. In this measurement the quasi-free $np$ reactions 
were selected by detection of fast spectator protons from the deuterium
breakup in the forward direction.

As has already been stated above, the HADES data on the dilepton invariant 
mass distribution (DIMD) in the $pp$ reaction were described well by the SM2 
model~\cite{aga09a,ram09} while the KK model overestimated them. However, 
for the quasi-free $np$ collisions, both models fail to reproduce the 
experimental cross sections. The shape of the DIMD changes dramatically 
when going from $pp$ to $pn$ interactions. One notices that (1) in the 
dilepton invariant mass region of 0.15 to 0.35 GeV, the yield is about 
9 times higher in the $np$ case as compared to that of the $pp$ case, 
(2) in the quasi-free $np$ case, the tail of the DIM extends to much 
larger values of $M$, (3) the ratio of quasi-free $np$ to $pp$ cross 
sections reaches almost a value of 100 at $M = 0.5 \,$ GeV.  
\begin{figure}
\epsfxsize=6cm
\begin{center}
\begin{tabular}{cc}
\epsfbox{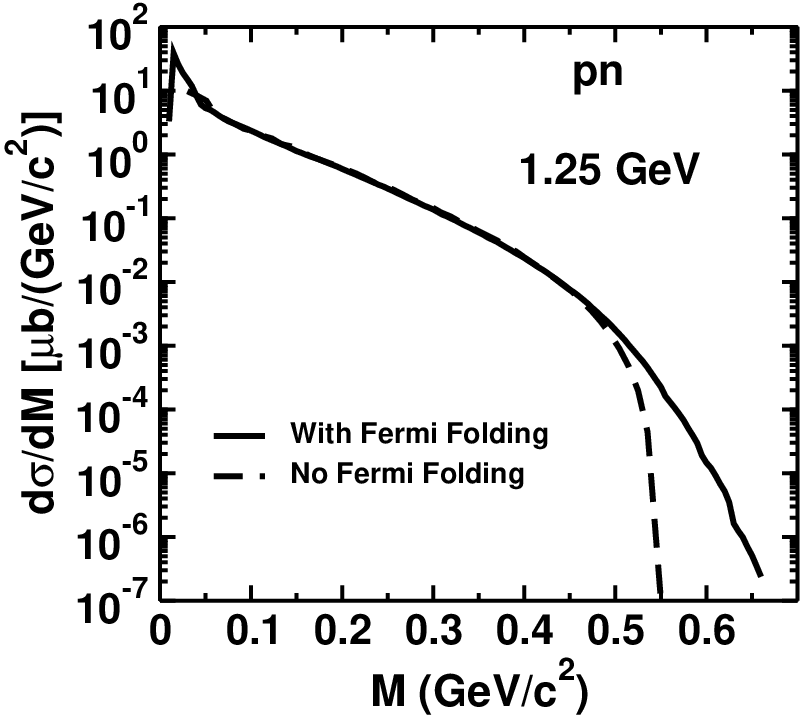} & \hspace{0.99cm}
\epsfxsize=6cm
\epsfbox{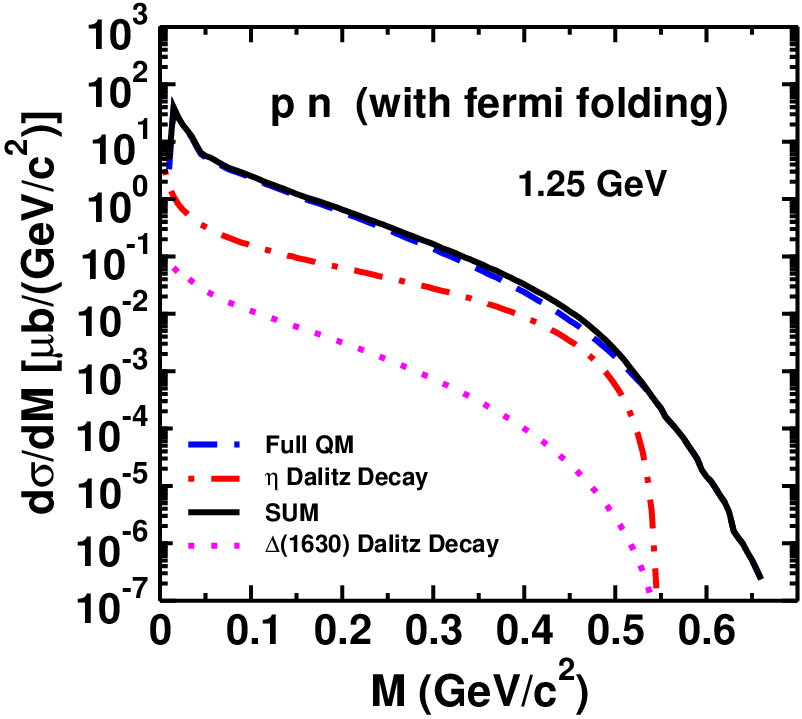}
\end{tabular}
\end{center}
\caption{(Left panel) the dilepton invariant mass distribution in $np$
(dashed line) and quasi-free $np$ interaction (solid line) at 1.25 GeV/nucleon. 
In the later case the center mass energy is smeared by including the neutron 
momentum distribution in the deuteron using the Argonne V18 potential.
(Right panel) the dilepton invariant mass distribution in the quasi-free 
$np$ collisions at 1.25 GeV/nucleon calculated within the model of 
Ref.~\protect\cite{shy09} (dashed line). Also shown are the contributions of
$\eta$ meson and $\Delta(1600)$ Dalitz decay processes by dashed-dotted and
dotted lines, respectively. Their incoherent sum is shown by the solid line. 
}
\label{fig:hades}
\end{figure}
\noindent

To understand points (1)-(3), we introduce the following 3 additional features
in the SM2 model for the $np$ case. (i) The available energy in the center of 
mass is smeared to include the neutron momentum distribution in the deuteron 
using the Argonne V18 potential~\cite{wir95}. The consequence of this 
procedure is that the $dp$ reaction results is a smeared $np$ reaction
with center of mass energies in excess of the threshold for the $\eta$ 
meson production (see, e.g., Ref.~\cite{ing09}). (ii) Hence, the contribution 
from the $\eta$ Dalitz decay is taken into account. In these calculations the 
total cross sections for the $n + p \to n + p + \eta$ reaction is taken from 
the Ref.~\cite{shy07} where a good description is achieved of the available 
corresponding experimental data. (iii) The dilepton yields from the Dalitz 
decay of higher mass delta resonances are included. Our preliminary results 
are shown in Fig.~5. It was already noted in Ref.~\cite{ram09} that the 
predictions of the SM2 model are in agreement with the $np$ data for
$M < 0.3$ GeV.  In the left panel of Fig. 5, we note that the Fermi folding
procedure leads to an extended tail in the invariant mass distribution.
In the right panel, we show the total cross sections obtained by Fermi folding
the cross sections of the SM2 model together with contributions of the
$\eta$ meson and $\Delta(1600)$ isobar Dalitz decay processes. Their 
incoherent sum is shown by the solid line. It is seen that the inclusion 
of contributions (ii) and (iii) leads to only a marginal enhancement in the 
summed cross sections over those of the SM2 model for $M$ values around 
$0.50$ GeV. Therefore, a proper understanding of point (3) is still lacking 
at this stage.  

\section{Summary and Conclusions}

In summary, we have shown that the dilepton yields from the $NN$ 
bremsstrahlung process calculated within the SM (1 $\&$ 2) models
\cite{shy03,shy09} are lower than those of the KK model~\cite{kap06,kap09} 
by factors of 2-4  for both $pp$ and $pn$ collisions at 1 - 2 GeV incident 
energies. The recent HADES measurements have confirmed that the cross sections 
calculated within the SM2 model are in good agreement with the experimental 
data for $pp$ reaction while those of the KK model overestimate them. 
This was also the case for the DLS data as shown in Fig. 3. This implies that 
care must be exercised in using the larger $NN$ bremsstrahlung cross sections 
in the transport model calculations in order to explain the DLS and HADES data 
obtained in the nucleus-nucleus collisions.

With HADES collaboration confirming the old DLS data, the resolution of the 
"DLS" puzzle has indeed now shifted to a better understanding of the theoretical 
models. In this context, the focus should be to achieve a proper explanation 
of the HADES data for dilepton production in quasi-free $np$ interaction.
We showed that smearing of the center of mass energy to include the neutron
momentum distribution in deuteron, does lead to extended tails in the dilepton 
invariant mass distributions. However, the absolute magnitudes of the 
cross sections in the region of dilepton invariant masses around 0.5 GeV are
still not explained even after taking into account the contributions
from the Dalitz decays of $\eta$ meson and higher mass resonances.    
Work is in progress to resolve this new puzzle.

\section{acknowledgments}

This work is supported by the Helmholtz International Center for FAIR 
(HIC for FAIR) under the LOWE program.


\begin{thebibliography}{99}
\bibitem{por97}
R. J. Porter {\it et al.}, Phys. Rev. Lett. {\bf 79}, 1229 (1997).
\bibitem{aga09}
G.Agakichiev {\it et al.}, Eur. Phys. J. {\bf A 41} (2009) 243.
\bibitem{aga05}
G. Agakichiev {\it et al.}, Eur. Phys. J. C {\bf 41}, 475 (2005).

\bibitem{ada03}
D. Adamova {\it et al.}, Phys. Rev. Lett. {\bf 91}, 042301 (2003).

\bibitem{arn06}
R. Arnaldi {\it et al.}, Phys. Rev. Lett. {\bf 96}, 162302 (2006).

\bibitem{afa07}
S. Afanasiev {\it et al.}, arXiv:0706.303 [nucl-ex].

\bibitem{rap07}
R. Rapp, J. Phys. G {\bf 34}, S405 (2007); R. Rapp, Nucl. Phys. {\bf A782},
275 (2007);
H. van Hees and R. Rapp, Phys. Rev. Lett. {\bf 97}, 102301 (2006);
J. Ruppert, C. Gale, T. Renk, P. Lichard, and J. I. Kapusta, Phys. Rev. Lett.
{\bf 100}, 162301 (2008).

\bibitem{ern98}
C. Ernst, S. A. Bass, M. Belkacem, H. St\"ocker, and W. Greiner, Phys. Rev. C
{\bf 58}, 447 (1998).

\bibitem{cas99}
W. Cassing and E. L. Bratkovskaya, Phys. Rep. {\bf 308}, 65 (1999).

\bibitem{rap00}
R. Rapp and J. Wambach, Adv. Nucl. Phys. {\bf 25}, 1 (2000).

\bibitem{she03}
K. Shekhter, C. Fuchs, A. Faessler, M. Krivoruchenko, and B. Martemyanov,
 Phys. Rev. C {\bf 68}, 014904 (2003).

\bibitem{aga07}
G. Agakichiev {\it et al.}, Phys. Rev. Lett. {\bf 98}, 052302 (2007).

\bibitem{bra98}
E. L. Bratkovskaya, W. Cassing, R. Rapp, and J. Wambach, Nucl. Phys.
{\bf A634}, 168 (1998).

\bibitem{cas98}
W. Cassing, E. L. Bratkovskaya, R. Rapp and J. Wambach, Phys. Rev. C {\bf 57},
916 (1998).

\bibitem{aga08}
G. Agakishiev {\it et al.}, Phys. Lett. {\bf B663}, 43 (2008).

\bibitem{ruc76}
R. R\"uckl, Phys. Lett. {\bf B64}, 39 (1976).

\bibitem{gal87}
C. Gale and J. Kapusta, Phys. Rev. C {\bf 35}, 2107 (1987);{\it ibid},
Phys. Rev. C {\bf 40}, 2397 (1989).

\bibitem{bra08}
E. L. Bratkovskaya and W. Cassing, Nucl. Phys. {\bf A807}, 214 (2008).

\bibitem{kap06}
L. Kaptari and B. K\"ampfer, Nucl. Phys. {\bf A764}, 338 (2006).

\bibitem{kap09}
L. Kaptari and B. K\"ampfer, Phys. Rev. C {\bf 80}, 064003 (2009)

\bibitem{sch94}
M. Sch\"afer, H. C. D\"onges, A. Engel and U. Mosel, Nucl. Phys. A 575, 429
(1994)

\bibitem{dej96}
F. de Jong and U. Mosel, Phys. Lett. {\bf B379}, 45 (1996).

\bibitem{shy03}
R. Shyam and U. Mosel, Phys. Rev. C {\bf 67}, 065202 (2003).

\bibitem{shy09}
R. Shyam and U. Mosel, Phys. Rev. C {\bf 79}, 035203 (2009).

\bibitem{shy98}
R. Shyam and U. Mosel, Phys. Lett. {\bf B426}, 1 (1998).

\bibitem{shy99}
R. Shyam, Phys. Rev. C {\bf 60}, 055213 (1999).

\bibitem{shy01}
R. Shyam , G. Penner and U. Mosel, Phys. Rev. C {\bf 63}, 022202(R) (2001).

\bibitem{shy07}
R. Shyam, Phys. Rev. C {\bf 75},055201 (2007).  
  
\bibitem{ris84}
D. O. Riska, Prog. Part. Nucl. Phys. {\bf 11}, 199 (1984).\

\bibitem{uso05}
A. Usov and O. Scholten, Phys. Rev. C {\bf 72}, 025205 (2005).

\bibitem{sch91}
M. Sch\"afer, T. S. Biro, W. Cassing, U. Mosel, H. Nifenecker and J. A.
Pinston, Z. Phys. A {\bf 339}, 391 (1991).

\bibitem{dej97}
F. de Jong and U. Mosel, Phys. Lett. {\bf B392}, 273 (1997).

\bibitem{aga09a}
G. Agakishiev {\it et al.}, arXiv:0910.5875 [nucl-ex].

\bibitem{ram09}
B. Ramstein {\it et al.}, arXiv:0912.2677 [nucl-ex].

\bibitem{wir95}
R. B. Wiringa, V. G. J. Stoks, and R. Schiavilla, 
Phys. Rev. C 51, 38 (1995).

\bibitem{ing09}
I. Fr\"ohlich {\it et al.}, arXiv:0909.5373
\end{thebibliography}
\end{document}